# Macroscopic quantum self-trapping and Josephson oscillations of exciton-polaritons

M. Abbarchi[1], A. Amo[1,*], V. G. Sala[1,2], D. D. Solnyshkov[3], H. Flayac[3], L. Ferrier[1], I. Sagnes[1], E. Galopin[1], A. Lemaître[1], G. Malpuech[3], J. Bloch[1]

[1] *CNRS-Laboratoire de Photonique et Nanostructures, Route de Nozay, 91460 Marcoussis, France*

[2] *Laboratoire Kastler Brossel, Université Pierre et Marie Curie, École Normale Supérieure et CNRS, UPMC Case 74, 4 place Jussieu, 75252 Paris Cedex 05, France*

[3] *Institut Pascal, PHOTON-N2, Clermont Université, University Blaise Pascal, CNRS, 24 avenue des Landais, 63177 Aubière Cedex, France*

* e-mail : alberto.amo@lpn.cnrs.fr

A textbook example of quantum mechanical effects is the coupling of two states through a tunnel barrier. In the case of macroscopic quantum states subject to interactions, the tunnel coupling gives rise to Josephson phenomena including Rabi oscillations, the a.c. and d.c. effects, or macroscopic self-trapping depending on whether tunnelling or interactions dominate. Non-linear Josephson physics, observed in superfluid helium and atomic condensates, has remained inaccessible in photonic systems due to the required effective photon-photon interactions. We report on the observation of non-linear Josephson oscillations of two coupled polariton condensates confined in a photonic molecule etched in a semiconductor microcavity. By varying both the distance between the micropillars forming the molecule and the condensate density in each micropillar, we control the ratio of coupling to interaction energy. At low densities we observe coherent oscillations of particles tunnelling between the two micropillars. At high densities, interactions quench the transfer of particles inducing the macroscopic self-trapping of the condensate in one of the micropillars. The finite lifetime of polaritons results in a dynamical transition from self-trapping to oscillations with π phase. Our results open the way to the experimental study of highly non-linear regimes in photonic systems, such as chaos or symmetry-breaking bifurcations.

A bosonic Josephson junction is a device in which two macroscopic ensembles of bosons, each of them occupying a single quantum state, are coupled by a tunnel barrier. The system can be described by the following coupled non-linear Schrödinger equations[1]:

$$i\hbar \frac{d\psi_L}{dt} = (E_L^0 + U|\psi_L|^2)\psi_L - J\psi_R, \qquad (1a)$$
$$i\hbar \frac{d\psi_R}{dt} = (E_R^0 + U|\psi_R|^2)\psi_R - J\psi_L, \qquad (1b)$$

where $\psi_{L,R}$ are the bosonic wavefunctions with particle densities $|\psi_{L,R}|^2$ localised to the left (L) and to the right (R) of the barrier, $E_{L,R}^0$ is the single particle energy of the quantum states, $U$ is the particle-particle interaction strength and $J$ is the tunnel coupling constant. In the absence of interactions, Eqs. 1a and 1b can be diagonalised in a basis of bonding ($2^{-1/2}[\psi_L + \psi_R]$) and antibonding states ($2^{-1/2}[\psi_L - \psi_R]$). An initial state prepared in a linear combination of these two (for instance, all particles in the left site) will result in density oscillations between the two sites. This is the main principle of the bosonic Josephson effect, which manifests in an ensemble of oscillatory regimes. In the absence of interactions, sinusoidal oscillations take place with a frequency $\hbar\omega = \sqrt{4J^2 + (E_L^0 - E_R^0)^2}$. This regime corresponds to the so-called Rabi oscillations and to the a.c. Josephson effect[2, 3].

Josephson physics shows the most spectacular phenomena in the non-linear regime, when the interaction energy ($U|\psi|^2$) is greater than the coupling $J$. The transfer of particles from one site to the other gives rise to a dynamical energy renormalisation resulting in non-harmonic oscillations. If interactions are strong enough ($U|\psi|^2 \gg J$), above an initial critical population imbalance between the two sites, the self-induced energy renormalisation quenches the tunnelling, and most of the particles remain localised in one of the sites. This out of equilibrium metastable regime is the so-called macroscopic quantum self-trapping.

A number of bosonic systems have demonstrated Josephson physics. Harmonic oscillations in the linear regime have been obtained in superconductor junctions or in nanoscale apertures connecting superfluid helium vessels[4, 5]. Bose-Einstein condensates of ultracold atoms in coupled traps have, additionally, shown macroscopic self-trapping[6], and collective phases in arrays of junctions[7]. For photonic systems, mostly theoretical results have been obtained so far. Despite the observation of symmetry breaking in photonic Kerr media[8, 9], non-linear Josephson oscillations have remained inaccessible due to the short lifetimes and very weak effective photon-photon interactions in standard non-linear optical systems.

In this work we present the experimental observation of the non-linear Josephson physics in a photonic system. We use condensates of microcavity polaritons in coupled micropillars etched out of a planar semiconductor microcavity[10]. The mixed light-matter nature of polaritons makes it possible on one hand to create a well-controlled double well by engineering their photonic component and, on the other hand, to boost the optical non-linearities thanks to the strongly interacting excitonic component. In our structures the micropillars partially overlap (see Fig. 1) creating a bosonic Josephson junction in which the tunnel coupling strength $J$ can be tuned by modifying the center-to-center separation[10]. Coherent oscillations in the linear regime between two coupled polariton condensates have been recently observed in a planar CdTe microcavity using two neighbouring photonic defects present in the random disorder of the structure[11]. Here we show the passage from coherent oscillations between the two sites with controlled frequency, to the extreme non-linear situation of trapping in a single micropillar induced by self-interactions [12, 13]. Since polariton condensates have a finite lifetime, the interaction energy changes dynamically and we observe in time the transition from the self-trapped to the oscillatory regime, and the onset of anharmonic phases.

The coupled micropillars used in our studies (Fig. 1a) are obtained by dry etching of a planar semiconductor microcavity with a Rabi splitting of 15 meV at 10 K, the temperature of our experiments (see Supplementary Information). Each individual pillar has a diameter of 4 μm and presents a series of confined polaritonic states[14-16]} with a lifetime of ~33 ps. The centre-to-centre separation of 3.73 μm results in a tunnel coupling of the confined polaritonic ground states of $J = 0.1$ meV. This double potential well system can be described by Eqs. 1a(b) with the addition of a phenomenological decay term[12, 17] $-i(\hbar/2\tau)\psi_{L(R)}$ accounting for the polariton losses due to the escape of photons out of the cavity, and with a positive value of $U$ coming from the polariton-polariton repulsive interactions[18].

The Madelung transformation $\psi_{L,R}(t) = \sqrt{N_{L,R}(t)}\ e^{i\theta_{L,R}(t)}$ allows us to rewrite Eqs. 1 in their dynamical form[2]:

$$\frac{\hbar}{2J}\dot{z} = \sqrt{1-z^2(t)}\sin\phi(t), \qquad (2a)$$

$$\frac{\hbar}{2J}\dot{\phi} = -\frac{E_L^0 - E_R^0}{2J} - \frac{UN_T\, e^{-t/\tau}}{2J}z(t) + \frac{z(t)}{\sqrt{1-z^2(t)}}\cos\phi(t), \qquad (2b)$$

where $z(t) = \frac{N_L - N_R}{N_T}$ is the population imbalance between the two micropillars (with $N_T = N_L + N_R$ the total population), and $\phi(t) = \theta_L(t) - \theta_R(t)$ is the phase difference. $E_L^0 - E_R^0$ is the energy difference between the ground states of left and right sites, intrinsically zero in our case as the two pillars are identical. The second term on the right hand side of Eq. 2b contains all the non-linear features of our system, and it is the only one that is affected by the decay (included phenomenologically via the parameter $\tau$, standing for the polariton lifetime).

In order to study the different Josephson regimes we excite the system with a 1.7 ps pulsed laser resonant with the ground state energy of the micropillars (~780 nm). The spectral width of the laser (~1.2 meV>$J$) allows us to initialize the system in a linear combination of bonding and antibonding states (B and AB in the figures)[19]. The exciton reservoir is not excited as it lays 7.5 meV above the ground state energy, and all observed non-linear effects arise from self-interactions within the polariton condensates. We use a 10 µm wide Gaussian excitation spot, covering the whole molecule. In this geometry, $\phi(0) = 0$ and the initial population imbalance $z(0)$ can be tuned by shifting the spot with respect to the centre of the molecule.

The light emitted from a single photonic molecule is collected with a microscope objective in reflection geometry and analysed in energy and time by means of a spectrometer coupled to a streak camera. In order to avoid the strong reflection of the laser beam in our detectors, we select the linear polarisation of emission that is perpendicular to that of the excitation (parallel to the molecule long axis). The molecules present an intrinsic linear polarisation splitting along a non-trivial direction which slowly rotates the polarisation direction of the injected polaritons[20, 21]. This allows us to measure in cross-polarised detection the energy, population imbalance $z(t)$ and phase difference $\phi(t)$ between the two sites (see Supplementary Information). The polarisation splitting is of the order of ~40 µeV, resulting in a polarisation rotation that is much slower (>70 ps) than the Josephson oscillation timescales in our system (up to 21 ps).

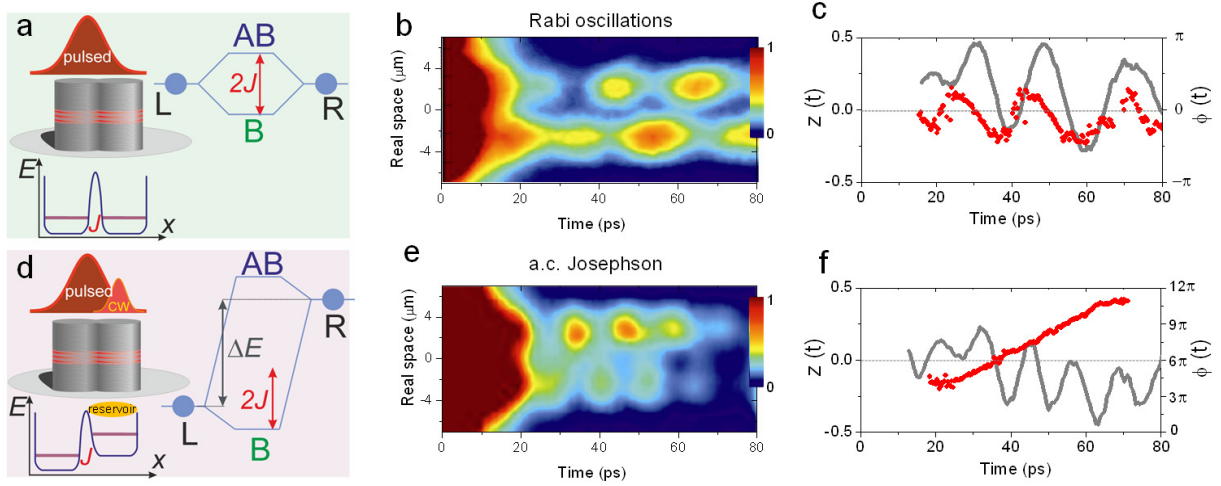

Fig. 1. **Rabi oscillations and a.c. Josephson effect. a** Scheme of a photonic molecule. The coupling *J* between the ground states of each micropillar (L, R) gives rise to bonding (B) and antibonding (AB) modes. **b** Emitted intensity when an off-centred Gaussian pulse at low power (2.5 mW) excites the system. **c** Measured population imbalance and phase difference, showing harmonic oscillations with a frequency given by 2*J* (0.2 meV). The addition of a cw beam on top of the right micropillar induces a static blueshift of its ground state energy as schematised in **d**. The larger bonding-antibonding splitting results in faster intensity oscillations as shown in **e**, and in a monotonously increasing phase difference, depicted in **f**.

*Linear regime: Rabi oscillations and a.c. Josephson effect.* At low excitation density ($2J \gg UN_t \approx 0$), interaction effects are negligible and the dynamics of the system is entirely dominated by the tunnel coupling. This is the situation presented in Fig. 1a-c: coherent oscillations of the population and the phase are observed when the system is prepared with an initial population imbalance of *z*(0)=0.45. The measured oscillation period of 21 ps coincides with that expected for the nominal coupling of *J*=0.1 meV in this molecule. The *z*(*t*) amplitude remains constant in time, as expected from Eq. 2b when $UN_T \to 0$. The periodic oscillations of both the population and the phase around zero correspond to the regime of Rabi oscillations of two coupled modes with $\Delta E \equiv E_L^0 - E_R^0 = 0$.

A different regime, characterised by a running phase, can be obtained inducing an energy splitting $E_L^0 - E_R^0 \gtrsim J$ between the ground states of the micropillars. To do so, we add a weak cw non-resonant beam (730 nm) focussed within one of the micropillars (the right one, see Fig. 1d-f). This beam creates an excitonic reservoir which interacts with the ground state polariton mode of that micropillar inducing a stationary and local energy blueshift of about 0.35 meV (see Suppl. Fig. 2). Nevertheless, particle self-interactions, (within the condensates), remain negligible ($UN_T \approx 0$). Figure 1e-f shows that the addition of this beam results in an acceleration of the oscillations, and in a phase difference $\phi(t)$ which monotonously increases with time. From Eq. 2a we see that a continuously growing phase results in oscillations of the populations (see section E of the Supplementary Information). This regime is analogous to the so-called a.c. Josephson effect, in which a constant voltage difference across two superconductors connected by a tunnel barrier gives rise to an oscillating current[22]. It corresponds to the recent observation of coherent oscillations between two localised states trapped within the photonic disorder of a CdTe microcavity[11].

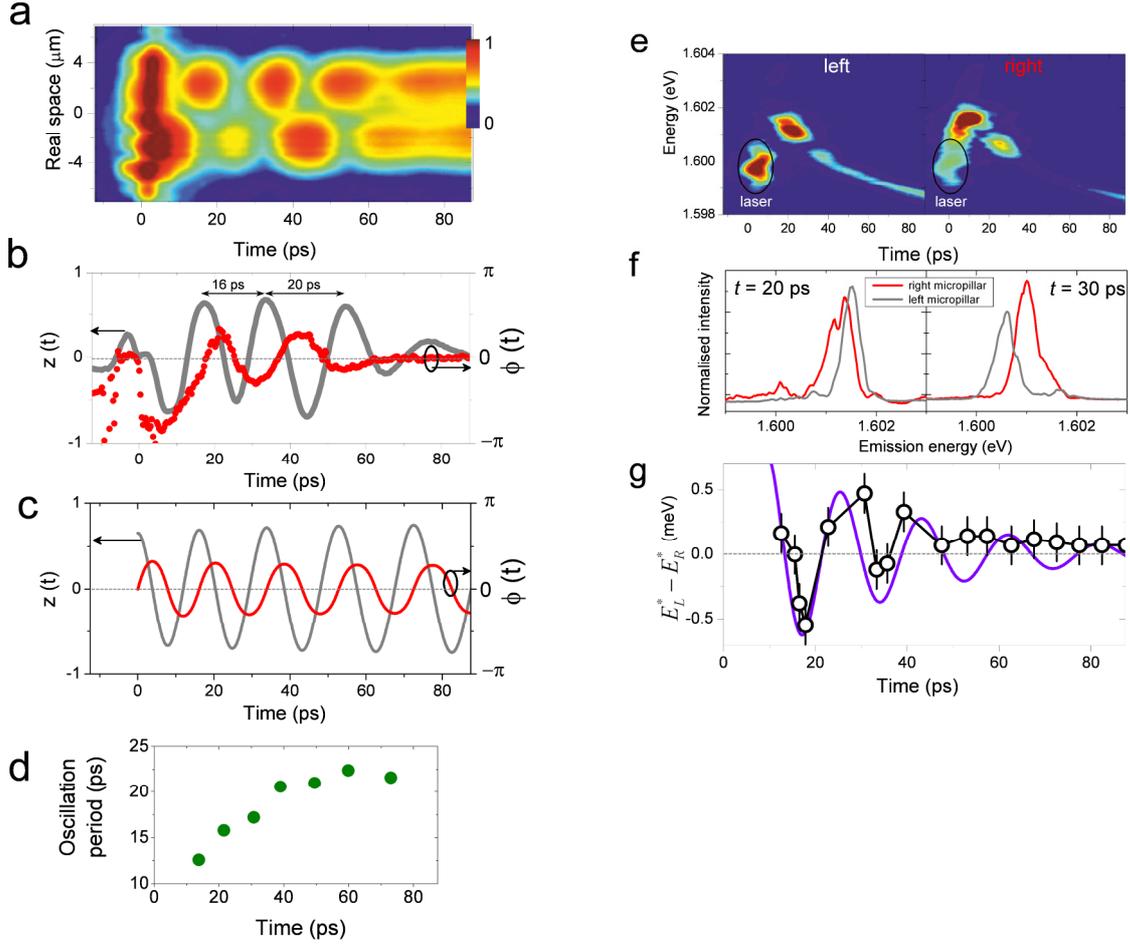

Fig. 2. **Anharmonic oscillations**. a Real space emission dynamics for a high power excitation (27 mW) and initial population imbalance of 0.62. **b** Shows the measured population imbalance along the phase difference, evidencing a monotonously increasing oscillation period, explicitly reported in **d**. **c** Displays the result of a simulation based on Eqs. 2 with $z(0) = 0.62$, $\phi(0) = 0$, $J$=0.1 meV, $UN_t(t=0)$= 1.2 meV and $\tau$=30 ps. **e** Energy- and time-resolved emission from each micropillar. The time dependent energy splitting can be extracted from the position of the emission of left and right micropillars at each time delay, as shown for $t$=20 and 30ps in **f**. **g** Energy difference between the ground state energy of each micropillar measured experimentally (dots) and simulated (solid line). The decaying oscillations arise from the dynamic energy renormalisation caused by the polariton-polariton interactions.

*Non-linear regime: anharmonic oscillations*. If the initial density is large enough ($UN_t \sim 2J$), interparticle repulsive interactions, characteristic of polariton condensates, strongly alter the dynamics. This is the situation shown in Fig. 2, in which a high density of polaritons is injected with an initial population imbalance $z(0) = 0.62$. The injected density in each micropillar is so large that the initial blueshift pushes the ground state energy of both micropillars above that of the excitation laser (Fig. 2e). We are thus in a regime of non-linear resonant absorption resulting in an interaction energy (proportional to $\sqrt{N_i}$ with $i$=L,R) much larger than the tunnel coupling $J$. In these conditions we expect a dynamical self-renormalisation of the energy levels in each site *i* when the high density of particles is transferred from one pillar to the other. We extract the renormalized energies in each pillar ($E_L^*, E_R^*$) from the energy resolved emission shown in Fig. 2e (see Supplementary Information), and we observe oscillations in the energy $E_L^*, E_R^*$ of each micropillar (Fig. 2f-g). This self-induced renormalisation results in a larger bonding-antibonding splitting and, therefore, in the decrease of the oscillation period with respect to the nominal tunnelling period $2\pi\hbar/2J$=21ps. This is observed at short times in Fig. 2a, b. At later times, the number of particles decreases, and so do the blueshifts in each site (Fig. 2e, g), eventually the oscillations tend to recover the harmonicity with a period given by *J* (Fig. 2d). The solution of Eqs. 2 given an initially high polariton density reproduces both the time evolution of $E_L^*, E_R^*$ and the anharmonicity of the oscillations, as shown in Fig. 2c and 2g.

*Macroscopic self-trapping.* In the high density regime, if the initial population imbalance is prepared above the threshold[2] $z(0) = 2J/UN_t$, a strongly asymmetric renormalisation of the energy takes place between the two sites (see Fig 3a). Under these conditions, the highly populated state is very close in energy and spatial distribution to the antibonding π-state. The system is thus in a quasi-metastable state in which most of the particles remain localised in one of the micropillars. This phenomenon, known as *macroscopic quantum self-trapping*, is depicted in Fig. 3b. At short times, the large energy difference between left and right micropillars results in a running phase (Fig. 3c,) similar to that of Fig. 1. In the present case, however, the energy difference arises entirely from self-interactions within the polariton condensates. At later times, the decay of particles reduces the energy splitting (Fig. 3d) and the oscillatory regime of the phase, dominated by the tunnel coupling, is recovered (Fig. 3c). Remarkably, when oscillations are set in, they take place around a value of the phase difference of $\pi$ (equivalently, an odd multiple of $\pi$), a situation also reproduced by simulations of Eqs. 2 (see Supplementary Information). The oscillations are probably damped due to intensity fluctuations of the pump laser, resulting in fluctuation in the recovery time of oscillations. Such $\pi$-oscillations appear naturally from the untrapping of the condensate from the quasi-antibonding mode, which is characterized by $\phi = \pi$. The dynamical transition from self-trapping to $\pi$-oscillations can be obtained only thanks to the finite particle lifetime and cannot be easily realised in systems such as Josephson junctions with atomic condensates.

In our experiments we have taken advantage of the large particle-particle interactions originating from the excitonic component of polaritons to demonstrate macroscopic self-trapping in a photonic system. This result opens the way to highly non-linear photonic phenomena like chaos [23-25], or spontaneous symmetry breaking and pitchfork bifurcations[8, 26-28] expected to give rise to highly squeezed macroscopic states[29, 30]. By reducing the size of the system, the many particle interactions evidenced here could be driven to the regime of single photon non-linearities[31], where phenomena such as photon blockade[32, 33], photon fermionisation[34] and the emergence of Bose-Hubbard physics[35-37] have been predicted.

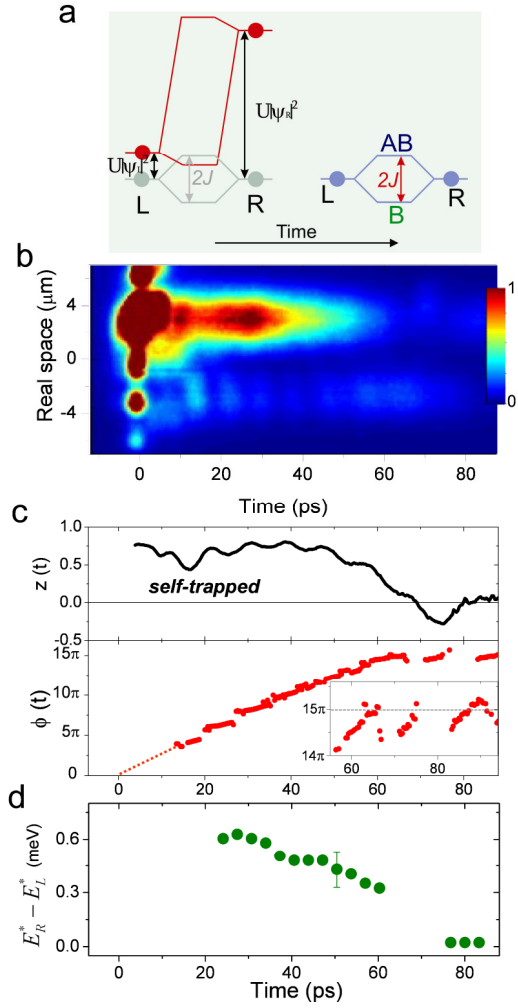

Fig. 3. **Macroscopic self-trapping**. **a** Scheme of the renormalized energy levels at short-time (self-trapped) and long-time delays (harmonic oscillations) for a highly asymmetric excitation at high density. **b** Measured dynamics of the emitted intensity. **c** Time evolution of the population imbalance and phase difference. At short times particles are self-trapped in the right micropillar, and the phase difference $\phi(t)$ increases linearly with time. At ~60 ps a transition to an oscillating regime takes place in the phase, induced by the escape of particles out of the microcavity. The red dotted line is an extrapolation of the phase evolution towards $t$=0. **d** Energy difference between the ground state of each micropillar extracted from the energy resolved dynamics (see Suplementary Material). The energy difference is induced by the asymmetric polariton density. The transition from self-trapping to oscillations takes place when $E_R^* - E_L^* \sim 2J$.

## Acknowledgements

We thank S. Barbay, A. Giacomotti, P.G. Kevrekidis and I. Zapata for fruitful discussions, P. Senellart for the lithography of the sample and T. Jacqmin for his experimental help. This work was supported by RENATECH, the Agence Nationale de la Recherche (contract "QUANDYDE"), the RTRA (contract "Boseflow1D"), the FP7 ITNs "Clermont4" (235114) and "Spin-Optronics"(237252), and the FP7 IRSES "Polaphen" (246912).

# Supplementary material for:

# Macroscopic quantum self-trapping and Josephson oscillations of exciton-polaritons

M. Abbarchi[1], A. Amo[1,*], V. G. Sala[1,2], D. D. Solnyshkov[3], H. Flayac[3], L. Ferrier[1], I. Sagnes[1], E. Galopin[1], A. Lemaître[1], G. Malpuech[3], J. Bloch[1]

[1] *CNRS-Laboratoire de Photonique et Nanostructures, Route de Nozay, 91460 Marcoussis, France*

[2] *Laboratoire Kastler Brossel, Université Pierre et Marie Curie, École Normale Supérieure et CNRS, UPMC Case 74, 4 place Jussieu, 75252 Paris Cedex 05, France*

[3] *Institut Pascal, PHOTON-N2, Clermont Université, University Blaise Pascal, CNRS, 24 avenue des Landais, 63177 Aubière Cedex, France*

* e-mail : alberto.amo@lpn.cnrs.fr

**A Experimental details**

*Sample and experimental set-up.* The planar sample was grown by molecular beam epitaxy and it consists of a λ/2 cavity sandwiched between two distributed Bragg reflectors containing 26 and 30 pairs, respectively, of $Al_{0.95}Ga_{0.01}As$, $Al_{0.20}Ga_{0.80}As$ λ/4 layers. The structure contains 3 groups of 4 GaAs quantum wells of 70Å width, located at the maxima of the electromagnetic field in the structure, resulting in a Rabi splitting of 15 meV. The quality factor of the etched structure is of 16000, resulting in a photon lifetime of 15 ps. As we perform the experiments in a photonic molecule with zero exciton-photon detuning, the polariton lifetime[10] is extended up to 30 ps.

Excitation of the sample is performed with a 1.7 ps pulsed laser resonant with the bonding and anti-bonding states. Its spectral width is of about 0.4 meV, which is larger than the bonding-antibonding splitting, thus allowing the preparation of a linear combination of both states. Simultaneously, it is smaller than the energy separation between the ground state and the first excited state in a single micropillar[15]. In our analysis we can thus disregard the excited states and only consider the ground state coupled modes.

The photoluminescence dynamics are recorded using a streak camera with a time resolution of 4 ps. The excitation pulse is linearly polarised along the long axis of the molecule, while detection is performed in the orthogonal polarisation.

*Measurement of the phase difference ϕ.* In order to measure the phase difference between the emissions from the two micropillars, we build a Michelson interferometer. An image of the emission of the molecule interferes with its mirror image at the entrance slit of a streak camera. In this way we interfere the emission of one of the micropillars of the molecule with that of the other micropillar.

Constructive/destructive interference appears depending on the phase difference between the emitted light from each micropillar, and also from the delay between the two arms of the interferometer. By varying the delay of one of the arms with a piezoelectric stage by 6π, we obtain an oscillating interference pattern from which we extract the phase difference $\phi(t)$ at different delay times[11] $t$.

*Energy resolved emission.* In order to obtain the energy of the emission, we image the micropillars on the entrance slit of a spectrometer placed in front of the streak camera. In this way we can measure the time evolution of the energy of the bonding ($E^*_B$) and antibonding ($E^*_{AB}$) states. From $E^*_{AB}$-$E^*_B$ we can calculate the energy difference between the renormalized left and right sites $E^*_L, E^*_R$ using the expression $E^*_L - E^*_R = \sqrt{(E^*_{AB} - E^*_B)^2 - 4J^2}$ (see Supplementary Material).

## B Energy resolved emission for the case of Rabi oscillations and a.c. Josephson effect.

Figure 1 shows the case in which polariton-polariton interactions play no role. In this case, the frequency of the oscillations is given by the splitting between the bonding and antibonding modes. We can access this splitting by performing an energy and time resolved experiment. Supplementary Fig. 1 shows the case of Rabi oscillations (Fig. 1a-c), in which the ground state of each micropillar has the same energy. By imaging the emission from one of the micropillars we measure a bonding-antibonding splitting of 0.2 meV, corresponding to twice the nominal coupling energy (0.1 meV).

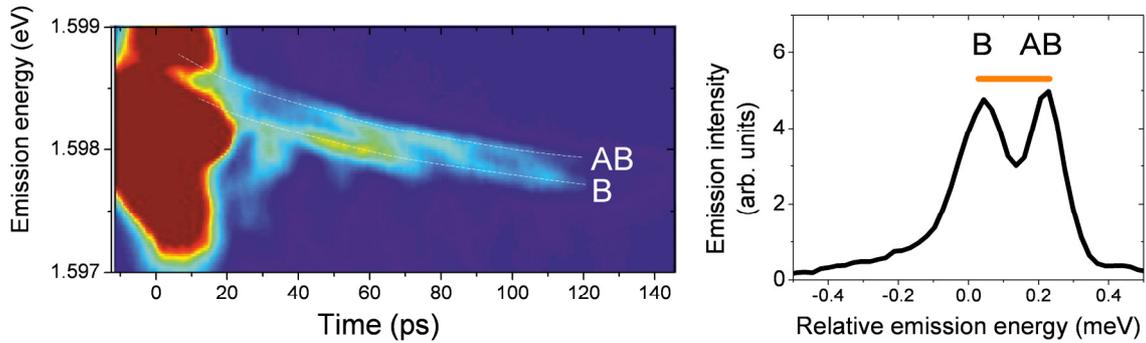

**Suppl. Fig. 1**. Energy resolved emission in the conditions of Fig. 1a-c (Rabi oscillations). The right panel shows a spectrum taken at 94 ps. A bonding-antibonding splitting of 0.2 meV is observed (orange bar).

In order to observe the a.c. Josephson effect, we shift the ground state energy of one of the micropillars (the right one, $E^0_R$) by adding a steady reservoir excitons with a cw laser beam. The bonding-antibonding splitting is given by the expression $E_{AB} - E_B = \sqrt{4J^2 + (E^0_L - E^0_R)^2}$ ). In this case $E^0_L \neq E^0_R$ and we expect a greater bonding-antibonding splitting as compared to that observed in Supplementary Figure 1. As depicted in Fig. 1e-f, oscillations are in this case asymmetric. The wavefunction of the antibonding state has a right pillar weight greater than the bonding state (whose center of gravity is shifted towards the left pillar). For this reason, the bonding-antibonding splitting is more clearly seen when looking at the energy resolved emission of both micropillars, as depicted in Supplementary Fig. 2: emission from the bonding state is more clearly seen in the left micropillar, and the antibonding emission is clearer in the right microcpillar. Note that the energy difference between bonding and antibonding states is constant all along the emission, evidencing negligible polariton interactions within the condensates.

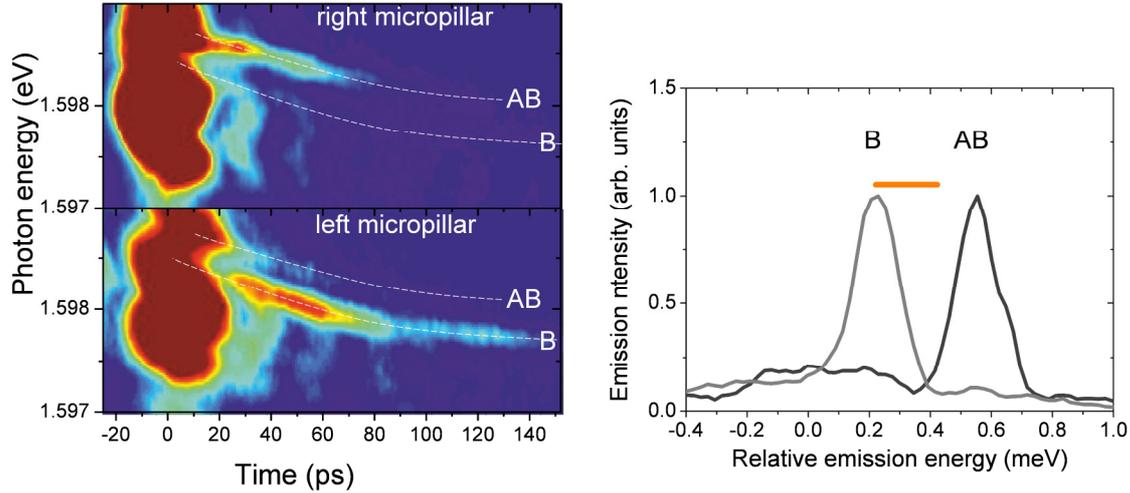

**Suppl. Fig. 2**. Energy resolved emission from the right and left micropillars in the conditions of Fig. 1d-f (a.c. Josephson effect). The left panel shows the spectra measured at a time delay of 72 ps for the left (grey) and right (black) micropillars. The orange bar indicates a splitting of 0.2 meV. A bonding-antibonding splitting of 0.38 meV is measured, from which we extract a $E_L^0 - E_R^0$ splitting of 0.35 meV.

### C Energy resolved emission in the self-trapped regime

We can use the procedure described above to measure the bonding-antibonding splitting in the self-trapped regime depicted in Fig. 3. Supplementary Figure 3 shows the energy and time resolved emission for this case. By measuring the bonding-antibonding splitting we can extract the energy difference of the renormalized left and right states. To do so, we use the same expression as in section C, by replacing $E_{L,R}^0$ by their renormalized values $E_{L,R}^*$: $E_{AB} - E_B = \sqrt{4J^2 + (E_L^* - E_R^*)^2}$. The energy difference calculated in this way is plotted in Fig. 3d.

The passage from the self-trapped to the oscillating regime that takes place at about t=60 ps (Fig. 3c), corresponds to a drop in $E_R^* - E_L^*$ below ~0.2 meV, which is twice the coupling energy $J$.

A similar procedure is used to extract the renormalized energy difference shown in Fig. 2 corresponding to the case of anharmonic oscillations.

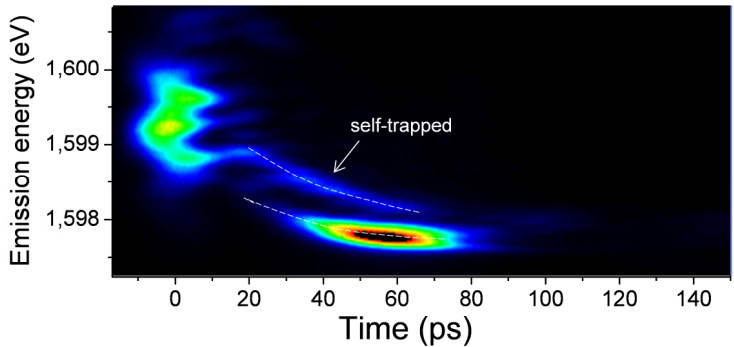

**Suppl. Fig. 3**. Energy resolved emission in the conditions of Fig. 3 (self-trapping) coming from the left micropillar (the low density one). At short time a large bonding-antibonding splitting is observed. At around ~60 ps, the splitting is reduced to 2$J$.

**D Polarisation rotation**

Our molecules present a linear polarisation splitting along a non-trivial direction (different from the crystallographic directions). Similar splittings, with a magnitude between 10-50 µeV were observed in GaAs planar microcavities via their effects on the pseudospin precession of propagating polaritons[20, 21]. The axis of the splitting depends on the exact position on the wafer and it is, thus, molecule dependent. In the photonic molecules we have studied, we inject polaritons with linear polarisation parallel to the molecule axis, which coincides with one of the highly symmetric crystallographic directions of the GaAs matrix. Detection is performed selecting a polarisation opposite to that of the excitation in order to avoid a strong stray laser light reflection in the detector. Initially, the injected polariton gas is polarised parallel to the excitation laser. The pseudospin precession induced by the linear polarisation splitting present in the molecule rotates the axis of polarisation allowing us to detect the emission in crossed polarisation.

We can directly observe the polarisation rotation by partially uncrossing the detection polariser. This is shown in Supplementary Figure 4 for the left pillar in the self-trapped case. The upper panel shows the emission with detection cross-polarised with respect to the linear polarisation of excitation. The maximum intensity is observed at a delay of 56 ps (red arrow). In co-polarised detection (lower pannel), the maximum of the emission takes place much earlier (red arrow). From these images we can extract a polarisation rotation period of ~100 ps corresponding to a polarisation splitting of ~40 µeV.

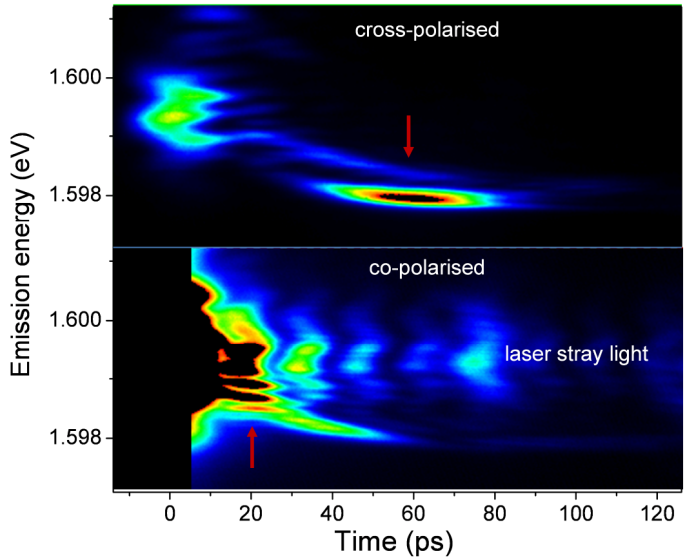

**Suppl. Fig. 4**. Energy resolved emission from the left pillar in the self-trapping conditions (Fig. 3). The upper panel shows the emission with detection cross-polarised with respect to the linear polarisation of excitation. The maximum intensity is observed at a delay of 56 ps (red arrow). In co-polarised detection (lower panel), the maximum of the emission takes place much earlier (red arrow), evidencing the rotation of the polarisation of emission caused by a polarisation splitting.

We can now provide an overall picture of the emission dynamics shown in Figs. 1-3. The time evolution of the emission reflects the interplay between the continuous decay of the density caused by the escape of photons out of the cavity (exponential decay with a lifetime of 30 ps), and the polarisation rotation with a period of ~100 ps. This interplay results in detected emission which has an almost constant intensity in the first ~100 ps. Indeed, at short times the number of particles in the system is very high but they do not have the time to significantly rotate their polarisation. At later times, the number of particles has decreased but their polarisation becomes parallel to that of the detection optics.

Note that the polarisation dynamics does not affect the population imbalance and phase difference dynamics studied in our work and described by Eqs. 1 and 2. The population imbalance measures the normalised difference in the number of particles in the two micropillars at each instant of time. It is thus not affected by the time evolution of the total emission. In the same way, the measured phase difference shows the instantaneous phase difference between left and right sites. In order to check this explicitly we have performed experiments with a detection of polarisation parallel to that of excitation. We observe the same phase difference and population imbalance dynamics as those reported in the main text in orthogonal polarisation. We do not show those data because the stray laser light (see lower panel in Supplementary Figure 4) does not provide images as clear and clean as the ones shown.

# E Simulations corresponding to Figs. 1 (Rabi oscillations and a.c. effect) and 3 (macrsocopic quantum self-trapping)

We can simulate the different observed oscillatory regimes by solving Eqs. 2 with the initial conditions used in our experiments.

In all our simulations we use a value of J=0.1 meV and an initial phase difference $\phi(0) = 0$. For the case of Rabi oscillations experimentally reported in Fig. 1a-c, we take a zero non-linearity ($UN_t = 0$), no splitting between the ground state energy of the two pillars $E_L^0 - E_R^0 = 0$, and an initial population imbalance $z(0)=0.45$. The result of the simulations is shown in Supplementary Figure 5a, with very good agreement with the experimental results.

We can simulate the a.c. effect reported in Fig. 1d-f with $E_L^0 - E_R^0 = 0.35$ meV and $z(0)=0.2$. The simulation is shown in Supplementary Figure 5b, in good agreement with the data.

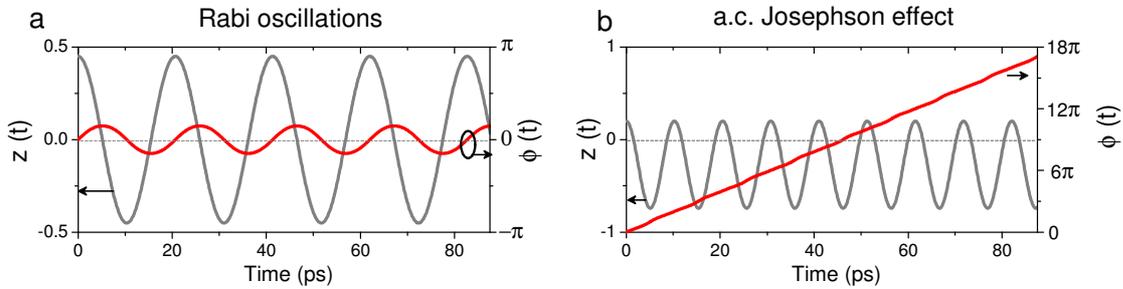

**Suppl. Fig. 5**. **a** Simulations of Eqs. 2 corresponding to the conditions of Fig. 1a-c (Rabi oscillations), with $z(0)=0.45$, $\phi(0) = 0$, $UN_t = 0$, $E_L^0 - E_R^0 = 0$. **b** Simulations corresponding to the a.c. Josephson effect (Fig. 1d-f), with $z(0)=0.2$, $\phi(0) = 0$, $UN_t = 0$, $E_L^0 - E_R^0 = 0.35$ meV.

We can simulate the self-trapping situation corresponding to the experiments shown in Fig. 3, by including a non-linearity via the second term of the right hand side of Eq. 2b. The simulation shown in Supplementary Figure 6 uses $UN_t = 5.5$ meV, $E_L^0 - E_R^0 = 0$, and $z(0)=0.98$, and $\tau$=30 ps, in good qualitative agreement with the results shown in Fig. 4. Note that we reproduce the phase oscillations around an odd multiple of π when coming out of the self-trapping regime. We have checked that this result is independent of the exact value of the lifetime: only oscillations of the π type appear after the self-trapping regime.

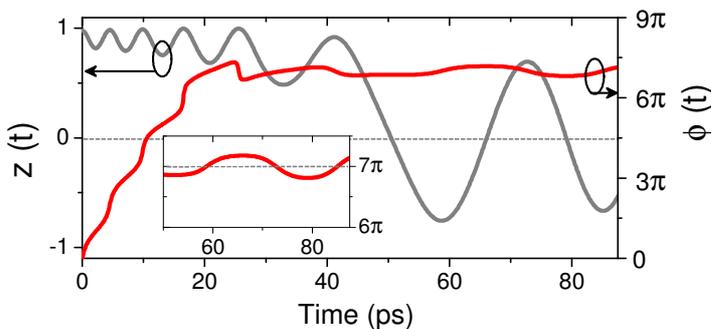

**Suppl. Fig. 6**. Simulations of Eqs. 2 corresponding to the conditions of Fig. 3 (macrsocopic quantum self-trapping), with $z(0)=0.95$, $\phi(0) = 0$, $UN_t = 5.5$ meV, $E_L^0 - E_R^0 = 0$.